# A Survey of Verification, Validation and Testing Solutions for Smart Contracts

Chaïmaa Benabbou
*Université Paris-Saclay, CEA, List*
F-91120, Palaiseau, France
chaimaa.benabbou@cea.fr

Önder Gürcan⋆
*Université Paris-Saclay, CEA, List*
F-91120, Palaiseau, France
onder.gurcan@cea.fr

*Abstract*—Smart contracts are programs stored on a blockchain that run when predetermined conditions are met. However, designing and implementing a smart contract is not trivial since upon deployment on a blockchain, it is no longer possible to modify it (neither for improving nor for bug fixing). It is only possible by deploying a new version of the smart contract which is costly (deployment cost for the new contract and destruction cost for the old contract). To this end, there are many solutions for testing the smart contracts before their deployment. Since realizing bug-free smart contracts increase the reliability, as well as reduce the cost, testing is an essential activity. In this paper, we group the existing solutions that attempt to tackle smart contract testing into following categories: public test networks, security analysis tools, blockchain emulators and blockchain simulators.Then, we analyze these solutions, categorize them and show what their pros and cons are.

## I. Introduction

With Bitcoin, the blockchain technology initially gained traction in 2008 [32]. Nowadays, it is considered as a major disruptive innovation with the potential to transform most industries. The blockchain can be regarded as a transparent, secure technology for storing and transmitting information that operates without a trusted third party.

For making the blockchain a general-purpose solution, Ethereum introduced the concept of smart contracts [39], *i.e.* immutable code on the blockchain that gets executed automatically once certain conditions are met between parties that do not necessarily trust each other. With smart contracts, the blockchain technology has seen a rapid climb to prominence, with its applications in various domains. They can improve insurance processes by automating claims when certain events occur, enable better supply chains[1] and more applications in business, commerce, and governance are still emerging . Given the wide range of smart contract adoption, best practices for implementing such code must be taken.

Since smart contracts deployment is definitive, the vulnerabilities and the attacks can challenge the sustainability of applications using them. The most known ones are related to Ethereum: the DAO attack in May 2016[2] that helped to gather around $150 million or the Parity Wallet hack[3], on November 8, 2017, that valued at over $152 million.

Moreover, executing smart contracts consumes certain amount of computation and memory. For instance, Ethereum community introduced *gas*, a notion to measure the amount of computational power needed to be paid to the miners in order to execute their operations.

Today, there are several blockchain systems that support smart contracts, such as Ethereum [39] and Quorum [5] that use Solidity for writing smart contracts, Hyperledger Fabric [2] that utilizes general-purpose programming languages (Go, Node.js, JavaScript and Java), Corda [8] that supports Java and Kotlin languages, Stellar [27] that uses Javascript SDK and Node.js and NEO [26] that develops smart contracts using C#. Besides, Tezos[15] designed new functional programming languages for smart contracts: Michelson, SmartPy and LIGO. Recently, Cardano[4] introduced smart contracts using Plutus, Marlowe & Glow programming languages. Besides, Tendermint/Cosmos[5] supports smart contracts written in any languages.

Bitcoin is also a pioneer in this field by enabling writing transactional rules using its non Turing-complete language, called Script[6] and soon a completely new way of designing smart contracts in Bitcoin will be introduced.

The life-cycle of a smart contract consists of the following stages [16, 19, 25, 36]:

- Analysis and design of the intended smart contract:
  - Define the requirements and the use-cases of the smart contract (*i.e.* the functions, the target blockchain, the target language and so on).
  - Design the clauses and the functions of the smart contract.
- Implementation and testing of the smart contract:

---

[1]Smart Contract and Supply Chain, https://www.frontiersin.org/articles/10.3389/fbloc.2020.535787/full, last access on 28/06/2021.
[2]DAO attack cost, https://blog.b9lab.com/the-dao-hack-in-eight-minutes-94919018692d, last access on 07/06/2021.
[3]Parity Wallet hack cost, https://cointelegraph.com/news/parity-multisig-wallet-hacked-or-how-come, last access on 07/06/2021.
[4]Cardano, https://docs.cardano.org/en/latest/index.html, last access on 16/06/2021.
[5]Tendermint, https://tendermint.com/core/, last access on 18/06/2021.
[6]Script, https://en.bitcoin.it/wiki/Script, last access on 15/06/2021.

- Select a testing approach and implement the smart contract using a corresponding language.
- Perform tests and analyze the results.

- Repetition of the above steps till the desired smart contract is implemented in the target language for the target blockchain.
- Deployment of the smart contract onto blockchain network definitively.
- Execution of the smart contract through functions calls:
  - Select a smart contract function and send a function call in the form of transaction.
  - Upon the confirmation of the transaction, the smart contract function is executed[7] and the blockchain state is updated[8].
- Termination of the smart contract to stop execution of its functions[9].

Consequently, implementation and testing phase is not trivial and should be taken into account seriously. To this end, there are various different verification, validation and testing (VV&T) solutions for different purposes proposed in the literature. However, there is no survey covering all the VV&T solutions for smart contracts.

Based on this observation, in this study, we provide a survey of these solutions. The survey is organized as follows. First, we divide the VV&T solutions into four distinctive categories and explain them (Section II). Then we analyze each category and compare them with respect to their proprieties (Section III). After, we provide a discussion about how these solutions can be used together (Section IV). Finally, we identify open challenges and conclude the paper (Section V).

## II. VV&T Solutions

From a software engineering point of view, to ensure the correctness of smart contracts, we use a technique conducted to perform verification, validation and testing (VV&T). Verification is the process to ensure that "we are developing the right product" (*i.e.* it meets the *specifications)*, validation is to ensure that "we have developed the product right" (*i.e.* it fulfills its intended *purpose*) and testing is to reveal "the existence of errors in the product", as stated in [6]. This involves verifying whether the detailed functional VV&T of business logic and process are operating as expected.

Since they are immutable code, once deployed, they are final and cannot be updated. The only way to fix a bug on a deployed smart contract is to deploy a new version of it; the old version remains there forever[10]. So performing VV&T before deployment make sure the smart contract has the expected behaviors.

To tackle these problems, many VV&T solutions have already been proposed and are still being proposed in the literature. To better analyze these solutions and compare them, we grouped them into four main categories : public test networks, security analysis tools, blockchain emulators with smart contracts support and blockchain simulators with smart contracts support.

### A. Public Test Networks

Blockchain systems are deployed onto public networks, called main networks. It is the end product available for the public to use. For VV&T purposes, there are also public available networks, called (public) test networks, in which anyone can access via their clients (*i.e.* popular wallet interfaces).

Public test networks enable to compile and deploy smart contracts and perform frequent tasks, such as running tests, automatically checking code for compilation errors or interacting with smart contracts in a public environment.

When using a public test network, the environment conditions are the same as the main network. The main difference is that, while in the main network crypto-currencies are traded for goods and services, in test networks they are given away for free without the need for mining. Hence, users do not have sufficient control to configure the network conditions. They are created and only configured by people from the community.

The main steps to test smart contracts using public test networks are:

- Choose a public test network depending on the desired target blockchain.
- Create a wallet to store the crypto-currencies using a blockchain client.
- Claim a desired amount of coins for free from a *Faucet Website*.
- Connect the blockchain client to the test network.
- Write a smart contract using a target language and deploy it onto the test network.
- For each smart contract function, write a test using a language that can interact with the blockchain client and execute it through the blockchain client.

There are several public test networks that exist in the literature. The Ethereum community proposed *Ropsten*, *Rinkby*, *Kovan* and *Görli* test networks[11]. While the Tezos community introduced Teztnets[12], a platform to help in the deployment of Tezos test networks, namely *Florencenet* and *Granadanet*. Testnet3[13] is another test network proposed by the Bitcoin community, nearly identical in characteristics to Bitcoin blockchain, that overcomes some difficulties related to transaction delay of the previous Testnet2. The Hyperledger Fabric community has also released two versions of *Fabric test network*[14] (v1.4 and v2.1) that enable developers to test their smart contracts and applications. Tendermint community provides gaia and basecoind tesnets for Tendermint/Cosmos-like blockchains.

---

[7] In fact, this is the case when the blockchain uses the Order-Execute-Update approach. However, there are also blockchains that use the Execute-Order-Validate-Update approach [2, 34].

[8] The world state can always be regenerated from the blockchain. It works as a cache that only stores the current value of a state.

[9] Here it should be noted that smart contract can not be deleted from the blockchain.

[10] Some permissioned blockchains like Hyperledger Fabric allows the smart contract logic to be updated so that the buggy version will not be accessible any longer.

[11] Ropsten Test Network: https://ropsten.etherscan.io/, Rinkby Test Network: https://rinkeby.etherscan.io/, Kovan Test Network: https://kovan.etherscan.io/ and Görli test network, https://goerli.net/ last accessed on 26/05/2021.

[12] Teztnets, https://teztnets.xyz/, last access on 04/06/2021.

[13] Bitcoin actual test network, https://en.bitcoinwiki.org/wiki/Testnet#Testnet_vs_Testnet2, last access on 14/06/2021.

*B. Security Analysis Tools*

Security analysis tools allow automatically analysing smart contracts and detecting some of their *predefined* common vulnerabilities. without *necessarily* deploying the smart contract to a blockchain test network.

To validate, verify and test smart contracts using such tools, it is necessary to write the smart contract and either invoke the dedicated tool command or add the tool plugin, if possible, to a blockchain emulator that supports it. The security analysis tools then analyze either the source code of the smart contract or its compiled blockchain virtual machine bytecode. Most of the security analysis tools in the literature are developed for the Ethereum blockchain [1, 3, 18]. The most well-known ones are *Why3* [33], *Oyente* [24], *Securify* [38], Osiris [11], Sereum [31], *Mythx*[15], *Mythril*[16], *SmartCheck*[17], *Manticore*[18], *MAIAN*[19], *Solgraph*[20], *Reguard* [23], Slither [14], *Fether* [40], *FSolidM* [28] , *VeriSolid* [29], *ZoKrates* [13] and *Vandal* [7].

Concerning other blockchain systems, to the best of our knowledge, there are only a few solutions, namely *Chaincode analyzer*[21] [21] and *Revive^^c*[22] for Hyperledger Blockchain, *Zeus* for Ethereum and Hyperledger [20], and *SODA* [10] for any blockchain that adopts EVM as its smart contract runtime.

*C. Blockchain Emulators with smart contract support*

Blockchain emulators are software programs that imitate the features of a blockchain. They enable reproducing blockchain networks locally to mimic the outer behavior of the main network features. In other words, they duplicate completely the main network in a virtual environment to emulate VV&T scenarios. Thus, it allows the developer to debug and test smart contracts.

Generally, the emulators come with a default network configuration file but it can be expanded to change the environment options (include more networks, choose the mining mode, adapt the block time and so on) or smart contracts options.

To test smart contracts on emulators, a user needs to:
- Configure the blockchain if wanted (most of emulators come with a default configurations).
- Write and compile a smart contract.
- Migrate (or deploy) the smart contract locally on to the emulated blockchain.
- Run automated tests or write quick and effective tests.

In the literature, several smart contract emulators exist: *Hyperledger Umbra*, *Hyperledger Caliper*, *Hardhat*, *Truffle*, *Brownie*, *Takamaka* [37], *Ganache*, *Blockbench* [12], *Hawk* [22], *Remix*, *Tenderly*, and *Embark*[23].

*D. Blockchain Simulators with smart contract support*

A blockchain simulator can mimic the internal behaviours of blockchains. They are used when there is the need for the blockchain to perform in an expected way. It helps developers create a copy of an existing blockchain into a virtual environment to get an idea about how, in our case, smart contracts work. It does not follow all the rules of a main network but simulate the relevant behaviors. It enables users design, implement and evaluate a blockchain with the desired configuration of the network's initial conditions using different settings and parameters.

To test smart contracts on simulators, a user should:
- Set up the blockchain configuration
- Write a smart contract
- Deploy the smart contract on the simulated blockchain
- Write tests for deployed smart contracts' functions

There are several blockchain simulators in the literature, yet, to the best of our knowledge, there is only one blockchain simulator supporting smart contracts: *Guantlet*[24] [4].

## III. ANALYSIS OF SOLUTIONS

VV&T solutions are used to build smart contracts more efficiently and aim at improving their security and correctness. In this section, we examine and compare the aforementioned VV&T solutions.

Table I summarizes the most known smart contracts vulnerabilities with respect to three scopes: (1) smart contract scope that occurs due to the source code (such as *integer overflow/underflow* when performing operations with value limitations), (2) application scope that represents interactions between smart contracts or smart contracts and other entities (such as *Front Running* when a user react to a transaction before it is included in the next block) and (3) blockchain system scope that exercises the full blockchain entities togheter

---

[14]Hyperledger Fabric test netxork , https://hyperledger-fabric.readthedocs.io/en/latest/test_network.html, last access on 26/05/2021.

[15]Mythx, https://mythx.io/about/,lastaccess26/05/2021, last access on 25/05/2021.

[16]Mythril, https://github.com/ConsenSys/mythril, last access on 25/05/2021.

[17]SmartCheck, https://smartcontracts.smartdec.net/, last access on 25/05/2021.

[18]Manticore, https://github.com/trailofbits/manticore, last access on 27/05/2021.

[19]MAIAN, https://github.com/ivicanikolicsg/MAIAN, last access on 27/05/2021.

[20]Solgraph, https://github.com/raineorshine/solgraph, last access on 27/05/2021.

[21]Chaincode analyzer, https://github.com/FujitsuLaboratories/ChaincodeAnalyzer, last access on 10/06/2021.

[22]Revive^^c tool, https://github.com/sivachokkapu/revive-cc, last access on 01/06/2021.

[23]Hyperledger Umbra: https://github.com/hyperledger-labs/umbra, Hyperledger Caliper: https://hyperledger.github.io/caliper/, Hardhat: https://hardhat.org/, Truffle Suite: https://www.trufflesuite.com/, Brownie framework: https://eth-brownie.readthedocs.io/en/stable/, Ganache: https://www.trufflesuite.com/docs/ganache/overview, Remix: https://remix.ethereum.org/, Tenderly: https://tenderly.co/transaction-simulator/, Embark: https://framework.embarklabs.io/docs/contracts_configuration.html last access on 09/06/2021.

[24]Guantlelt, https://gauntlet.network/, last access on 16/06/2021.

(such as *Transaction Order Dependence* when the order of transactions can be easily manipulated) .

Table III summarizes the available VV&T solutions for smart contracts w.r.t target blockchains[25], smart contract languages, testing languages, vulnerabilities and Table II presents the level of control a developer have while using these different categories. We can easily see that the security analysis tools available are numerous compared to other categories. More precisely, we have 10 public test networks, 20 security analysis tools, 10 blockchain emulators with smart contract support and only 1 blockchain simulator with smart contract support.

Besides, it is also easy to see the vast majority of solutions support Ethereum while support for other blockchains is more scarce. Bitcoin, despite being the most known blockchain, has only one VV&T solution, a public test network that supports smart contracts (i.e. scripts), but there is no tool or emulator available for it.

## A. Target Blockchain

As shown in Table III, most of the VV&T solutions are dedicated to a very limited number of target blockchains. It is natural that, all public test networks are dedicated to specific target blockchains. Each target blockchain has at least one dedicated public test network. Most security analysis tools, however, including Ropsten, Rinkby, Oyente, Mythril and Smartcheck support only the Ethereum blockchain. There are only two tools supporting Hyperledger. Concerning Tendermint/Cosmos-like blockchains, most security analysis tools are supported, while no tools are available for the other target blockchains.

Blockchain emulators are similar to public test networks and thus they are usually dedicated to a single target blockchain or its derivations. Hardhat, Truffle and Brownie, for instance, support Ethereum and its variations. But, blockchain emulators also sometimes support other blockchains. Hyperledger Caliper and Blockbench, for example, support Ethereum and derivations of Hyperledger.

There is only one simulator solution but it is better at supporting various target blockchains if they are well designed. The simulator identified, Gauntlet, supports several types of target blockchains, namely Ethereum and Tendermint/Cosmos.

## B. Smart Contract and Test Languages

Concerning programming languages, most VV&T solutions support different ones for writing smart contracts and writing tests (Table III). The test networks for Ethereum support Solidity for smart contracts while tests are written in Javascript. Hyperledger Fabric and Tendermint/Cosmos test networks provide a larger range of languages, such as Solidity, Go, Node.js, Java, Javascript and Python for both smart contracts and tests.

When we look at security analysis tools, since most of them are dedicated to Ethereum, they are mostly supporting Solidity as a smart contract language. Most of these tools do not permit testers to implement their tests and provide predefined commands. Concerning emulators, they are similar to test networks and they support dedicated languages of the target blockchains. Concerning simulators, Gauntlet supports python as programming language.

| Vulnerab. & Attacks \ Scope | Smart Contract | Application | Blockchain System |
|---|---|---|---|
| Integer Overflow/Underflow | ✓ | X | X |
| Call to the unknown | X | X | ✓ |
| tx.origin usage | X | ✓ | X |
| Integer Bugs | ✓ | X | X |
| Callstack depth exception | X | ✓ | X |
| Unchecked Send Bug | ✓ | X | X |
| Mishandled Exceptions | ✓ | X | X |
| Reentrency | X | ✓ | X |
| Front Running | X | ✓ | X |
| Timestamp Dependence | X | ✓ | X |
| Black listed imports | X | X | ✓ |
| Transaction Order Depend. | X | X | ✓ |
| Global State Variables | X | X | ✓ |
| Gas costly patterns | ✓ | X | X |
| Read after write | ✓ | X | X |
| DoS Attack | X | ✓ | X |
| DAO attack | X | ✓ | X |

TABLE I: Scopes of vulnerabilities and attacks [21, 30]

## C. Vulnerabilities

Among all VV&T solutions, only security analysis tools come with pre-defined vulnerability tests that can be directly used (Table III). More in detail, all security analysis tools have pre-defined vulnerabilities and it is also possible to improve the supported vulnerabilities using plug-ins (e.g., FSolidM and VeriSolid). Public test networks and simulators, in contrast, provide manual VV&T by user-defined scenarios. Emulators provide manual and automated VV&T since they can add security analysis tools as plug-ins. For example, Remix and Truffle can add Mythx as a plug-ins and make use of its pre-defined vulnerabilities.

Regarding the vulnerabilities, the ones that are addressed the most by security analysis tools are Reentrancy, Timestamp dependence, Transaction Ordering Dependence and Integer overflow/underflow as shown in Table III.

Considering Table I, other vulnerabilities[26] exist with respect to different scopes. Security analysis tools mostly focus on the smart contract code and can not perform VV&T from the blockchain system scope as they do not have a global view of it. Considering other approaches, they enable us checking for different scopes of vulnerabilities.

## D. Parameter Control

In Table II, a comparison is made between the four categories considering the configuration parameters in control. For that, we fix a set of network, blockchain and smart contract parameters:

- Network-level paramaters:

---

[25]"Target blockchain" covers all the blockchains based on the principle one (*e.g.,* Ethereum covers Parity, Quorum and so on.).

[26]Smart Contract Weakness Classification Registry, https://swcregistry.io/, last access on 01/07/2021

| **VV&T Solutions** | | | Test Net. | S.A. Tools | Bc. Emu | Bc. Simu |
|---|---|---|---|---|---|---|
| **Parameters** | *Network* | Size | x | x | Partial | ✓ |
| | | Msg trans. delay | x | x | Partial | ✓ |
| | | Msg trans. reliability | x | x | Partial | ✓ |
| | *Blockchain* | Genesis Block | ✓ | Partial | ✓ | ✓ |
| | | Tx fees | x | x | Partial | ✓ |
| | | Block size | x | x | Partial | ✓ |
| | | Batch time out | x | x | Partial | ✓ |
| | | BlockCreation Mode | x | x | Partial | ✓ |
| | *Smart Contract* | Size | x | x | Partial | ✓ |
| | | Dependencies | ✓ | ✓ | ✓ | ✓ |

TABLE II: Control of parameters

- Size : number of nodes of the network
- Message transmission delay : communication delay between nodes
- Message transmission reliability : the ability of a message to be successfully transmitted within its deadline.
- Blockchain-level parameters:
  - Tx Fees : transaction fees
  - BlockSize : the maximum size of blocks
  - BatchTimeOut : Block creation frequency.
  - BlockCreationMode : define the mining mode (e.g., mine when needed, always mine, turn off the miner), the consensus to consider, its difficulty,
- Smart contract level parameters:
  - size : the maximum size of a smart contract
  - dependencies : specifies the smart contracts it depends on.

In public test networks and blockchain emulators, blockchain-level parameters (e.g., chain configuration, level of difficulty to mine blocks) are defined in the genesis block. In blockchain simulators, on the other hand, the parameter are usually configured in a configuration file which is used by the simulator to initialize the simulation.

## IV. DISCUSSION

Despite the advances in smart contracts VV&T solutions during the last couple of years, our study highlights several open challenges to be tackled by future work. We identify the following challenges:

### A. Pre-defined vs User-defined vulnerabilities

Most security analysis tools have a useful set of pre-defined vulnerabilities to analyze smart contracts and thus save time of developing user-defined scenarios. However, these tools consider generic properties and do not capture specific vulnerabilities such as the *Front Running* vulnerability which is one of the major security issues. Thus, these tools are effective in detecting typical errors but ineffective in detecting atypical vulnerabilities.

Therefore, manual generation of meaningful scenarios becomes a necessity. It enables us inspect deep corner cases that users are already aware of, given the wide range of vulnerabilities that can occur in smart contracts. Still, developing them could be very costly in time.

The appropriate way to proceed is a mix between pre-defined and user-defined solutions, which will speed up the VV&T process. In that way, we do not waste time in writing scenarios that security analysis tools can already detect and only focus on detecting the non-defined ones.

### B. Programming Languages

As demonstrated in Table III, most testing solutions support a domain-specific language (DSL), Solidity, as a smart contract language. Others (most recent ones) use common general-purpose languages such as Java that were not initially conceived for smart contracts and thus make less expressive but easier to developers since they are more familiar with them. Hence, the vulnerabilities and risks associated to a certain language are not the same with others, due to each language features and restrictions. With DSLs, very useful libraries that facilitate the programming of smart contracts are provided. While with general-purpose languages, there are no features or restrictions to write safe smart contracts. For example, using Go language, a non determinism may arise, leading to random number generation or global variables vulnerabilities. This does not mean that with general-purpose languages, more vulnerabilities will occur but using the available libraries in DSLs (*i.e* Solidity provides reusable behaviors and implementations of various standards such as OpenZeppelin, HQ20, DappSys), a consequent number of them can be avoided due to specific restrictions.

### C. Community Participation

A smart contract project can be more successful if a community of volunteers who are globally distributed participate to the VV&T process through the Internet, as also identified by the open source community [35]. This enables the community to make more tests and thus helps in detecting a large range of vulnerabilities. There are two ways to this: (1) the smart contract can be deployed onto a public test network and thus made available to the community, or (2) the smart contract project can be made available to community through a public repository.

### D. Confidentiality

There is a big interest from industries to adapt their existing services or new services as smart contracts. To this end, they are actively doing several proof-of-concept studies. During these studies, the industries usually use confidential information (e.g. sensitive data, confidential algorithms, solutions) that they do not want them to be shared publicly until they finalize their smart contract based solution. In such cases, obviously public test networks are not suitable as a VV&T approach. It is rather better using the other three approaches (i.e. security analysis tools, blockchain emulators or blockchain simulators) without putting their smart contract in a public environment. The disadvantage of this is that they cannot benefit from the community involvement that can speed up vulnerability exploration.

|  | Target Blockchain | | | | | Smart Contract Language | | | | | | Test Language | | | | | | Vulnerabilities | | | | | |
|  | | | | | | | | | | | | | | | | | | Pre-defined | | | | | |
|  | Bitcoin | Ethereum | Hyperledger Fabric | Tendermint/Cosmos | Others | (Bitcoin) Script | Solidity | Go | Java | JavaScript | Others | Solidity | Go | Java | JavaScript | Python | Others | Reentrency | Trans. order depen. | Timestamp depen. | Int. over-/under-flow | Others | User-defined |
| **Public Test Networks** | | | | | | | | | | | | | | | | | | | | | | | |
| Ropsten | x | ✓ | x | x | x | x | ✓ | x | x | x | x | x | x | x | ✓ | x | x | x | x | x | x | x | ✓ |
| Rinkby | x | ✓ | x | x | x | x | ✓ | x | x | x | x | x | x | x | ✓ | x | x | x | x | x | x | x | ✓ |
| Görli | x | ✓ | x | x | x | x | ✓ | x | x | x | x | x | x | x | ✓ | x | x | x | x | x | x | x | ✓ |
| Testnet3 | ✓ | x | x | x | x | ✓ | x | x | x | x | x | x | x | x | x | x | ✓ | x | x | x | x | x | ✓ |
| Florencenet | x | x | x | x | ✓ | x | x | x | x | x | ✓ | x | x | x | x | x | ✓ | x | x | x | x | x | ✓ |
| Granadanet | x | x | x | x | ✓ | x | x | x | x | x | ✓ | x | x | x | x | x | ✓ | x | x | x | x | x | ✓ |
| Cardano Testnet | x | x | x | x | ✓ | x | x | x | x | x | ✓ | x | x | x | x | x | ✓ | x | x | x | x | x | ✓ |
| Hyperledger Testnet | x | x | ✓ | x | x | x | x | ✓ | ✓ | ✓ | x | x | ✓ | ✓ | ✓ | x | ✓ | x | x | x | x | x | ✓ |
| Gaia | x | x | x | ✓ | x | x | ✓ | ✓ | ✓ | ✓ | x | x | ✓ | ✓ | ✓ | x | ✓ | x | x | x | x | x | ✓ |
| Basecoind | x | x | x | ✓ | x | x | ✓ | ✓ | ✓ | ✓ | x | x | ✓ | ✓ | ✓ | x | ✓ | x | x | x | x | x | ✓ |
| Total | 1 | 3 | 1 | 2 | 3 | 1 | 5 | 3 | 3 | 3 | 6 | 0 | 3 | 3 | 6 | 0 | 7 | 0 | 0 | 0 | 0 | 0 | 10 |
| **Security Analysis Tools** | | | | | | | | | | | | | | | | | | | | | | | |
| Oyente | x | ✓ | x | ✓ | x | x | ✓ | x | x | x | x | x | x | x | x | x | x | ✓ | ✓ | ✓ | x | ✓ | x |
| Mythx | x | ✓ | x | ✓ | x | x | ✓ | x | x | x | x | x | x | x | x | x | x | ✓ | ✓ | ✓ | ✓ | ✓ | x |
| Mythril | x | ✓ | x | ✓ | x | x | ✓ | x | x | x | x | x | x | x | x | x | x | ✓ | ✓ | ✓ | ✓ | ✓ | x |
| SmartCheck | x | ✓ | x | ✓ | x | x | ✓ | x | x | x | x | x | x | x | x | x | x | ✓ | ✓ | ✓ | ✓ | ✓ | x |
| Securify | x | ✓ | x | ✓ | x | x | ✓ | x | x | x | x | x | x | x | x | x | x | ✓ | ✓ | x | x | ✓ | x |
| Osiris | x | ✓ | x | ✓ | x | x | ✓ | x | x | x | x | x | x | x | x | x | x | x | x | x | x | ✓ | x |
| Sereum | x | ✓ | x | ✓ | x | x | ✓ | x | x | x | x | x | x | x | x | x | x | ✓ | x | x | x | x | x |
| Manticore | x | ✓ | x | ✓ | x | x | ✓ | x | x | x | x | x | x | x | x | x | x | ✓ | x | ✓ | ✓ | ✓ | x |
| MAIAN | x | ✓ | x | ✓ | x | x | ✓ | x | x | x | x | x | x | x | x | x | x | x | x | x | x | ✓ | x |
| Solgraph | x | ✓ | x | ✓ | x | x | ✓ | x | x | x | x | x | x | x | x | x | x | x | x | x | x | ✓ | x |
| Reguard | x | ✓ | x | ✓ | x | x | ✓ | x | x | x | x | x | x | x | x | x | x | ✓ | x | ✓ | x | ✓ | x |
| Slither | x | ✓ | x | ✓ | x | x | ✓ | x | x | x | x | x | x | x | x | x | x | ✓ | x | ✓ | x | x | x |
| Fether | x | ✓ | x | ✓ | x | x | ✓ | x | x | x | x | x | x | x | x | x | x | x | x | x | x | x | x |
| FSolidM | x | ✓ | x | ✓ | x | x | ✓ | x | x | x | x | x | x | x | x | x | x | x | x | x | x | ✓ | ✓ |
| VeriSolid | x | ✓ | x | ✓ | x | x | ✓ | x | x | x | x | x | x | x | x | x | x | x | x | x | x | ✓ | ✓ |
| Zokrates | x | ✓ | x | ✓ | x | x | ✓ | x | x | x | x | x | x | x | x | x | x | x | x | x | x | ✓ | x |
| Vandal | x | ✓ | x | ✓ | x | x | ✓ | x | x | x | x | ✓ | x | x | x | x | x | ✓ | x | x | x | ✓ | x |
| Chaincode Analyzer | x | x | ✓ | x | x | x | x | ✓ | x | x | x | x | ✓ | x | x | x | x | x | x | ✓ | x | ✓ | x |
| Zeus | x | ✓ | ✓ | ✓ | x | x | ✓ | ✓ | ✓ | ✓ | x | x | x | x | x | x | ✓ | ✓ | ✓ | x | ✓ | x | x |
| SODA | x | ✓ | x | ✓ | x | x | ✓ | x | x | x | x | x | ✓ | x | x | x | x | ✓ | x | ✓ | x | ✓ | x |
| Why3 | x | ✓ | x | x | x | x | x | x | x | x | ✓ | x | x | x | x | x | ✓ | x | x | x | x | x | ✓ |
| Total | 0 | 19 | 2 | 19 | 0 | 0 | 19 | 2 | 1 | 1 | 0 | 1 | 2 | 0 | 0 | 0 | 1 | 10 | 5 | 8 | 5 | 18 | 0 |
| **Blockchain Emulators with smart contract support** | | | | | | | | | | | | | | | | | | | | | | | |
| Hyperledger Umbra | x | x | ✓ | x | x | x | ✓ | x | x | x | x | x | x | x | ✓ | x | x | x | x | x | x | x | ✓ |
| Hyperldeger Caliper | x | ✓ | ✓ | x | ✓ | x | ✓ | x | ✓ | ✓ | x | x | ✓ | ✓ | ✓ | x | x | x | x | x | x | x | ✓ |
| Hardhat | x | ✓ | x | x | x | x | ✓ | x | x | x | x | ✓ | x | x | ✓ | x | x | x | x | x | x | x | ✓ |
| Truffle | x | ✓ | x | x | x | x | ✓ | x | x | x | x | ✓ | x | x | ✓ | x | ✓ | x | x | x | x | x | ✓ |
| Brownie | x | ✓ | x | x | x | x | ✓ | x | x | x | x | x | x | x | x | x | ✓ | x | x | x | x | x | ✓ |
| Ganache | x | ✓ | x | x | x | x | ✓ | x | x | x | x | x | x | x | ✓ | x | x | x | x | x | x | x | ✓ |
| Blockbench | x | ✓ | ✓ | x | x | x | x | ✓ | ✓ | ✓ | ✓ | x | ✓ | ✓ | ✓ | x | x | x | x | x | x | x | ✓ |
| Remix | x | ✓ | x | x | x | x | ✓ | x | x | x | x | ✓ | x | x | x | x | x | x | x | x | x | x | ✓ |
| Tenderly | x | ✓ | x | x | x | x | x | x | x | ✓ | x | x | x | x | ✓ | x | x | x | x | x | x | x | ✓ |
| Embark | x | ✓ | x | x | x | x | ✓ | x | x | ✓ | x | x | x | x | ✓ | x | x | x | x | x | x | x | ✓ |
| Total | 0 | 9 | 3 | 0 | 1 | 0 | 6 | 2 | 2 | 4 | 1 | 3 | 2 | 2 | 7 | 1 | 2 | 0 | 0 | 0 | 0 | 0 | 10 |
| **Blockchain Simulators with smart contract support** | | | | | | | | | | | | | | | | | | | | | | | |
| Gauntlet | x | ✓ | x | ✓ | ✓ | x | x | x | x | x | ✓ | x | x | x | ✓ | x | x | x | x | x | x | x | ✓ |
| Total | 0 | 1 | 0 | 1 | 1 | 0 | 0 | 0 | 0 | 0 | 2 | 0 | 0 | 0 | 1 | 0 | 0 | 0 | 0 | 0 | 0 | 0 | 2 |
| **Total** | 1 | 32 | 6 | 22 | 5 | 1 | 30 | 7 | 6 | 8 | 9 | 4 | 7 | 5 | 13 | 2 | 10 | 10 | 5 | 5 | 4 | 18 | 22 |

TABLE III: Resume of VV&T solutions wrt the target blockchain, smart contract language programming, test language programming and type of vulnerabilities detected.

*E. Flexibility of parameters*

An important point to mention about these testing solutions is the network size. Actually, testing solutions have generally fewer nodes than a main network. For example, the size of the main Bitcoin blockchain on May 20, 2021 is about 337

Gigabytes[27] whereas the size of the Testnet3 blockchain in 2020 is only about 32 Gigabytes. This makes transactions lighter and faster which simplifies performing attacks as it does not require large capacity. Also, depending on the control parameters, we could perform different tests, whether on vulnerabilities, latency, performance and so on.

*F. Levels of testing*

In traditional testing literature, the major levels of testing are unit-, integration- and system-levels, and some type of acceptance-level [9]. Basically, unit testing is used to make sure that the implemented code meets the user specifications and works properly. Integration testing, however, is used to combine different parts and test them jointly (in our case, it could be different smart contracts, or the smart contract and the user interactions). Finally, system testing is used to test all the components to ensure the overall right functioning.

Considering smart contracts, the scopes of potential vulnerabilities are identified in Table I. Based on this observation, we can say that smart contract, decentralized application and blockchain system level vulnerabilities should be tested using unit, integration and system level testing respectively.

As shown in Table III, different testing solutions provides mechanisms for testing smart contracts against different types of vulnerabilities and attacks. It is up to the developer/tester to identify the tests to be accomplished and to choose the appropriate testing solutions.

For instance, as shown in Section III, except for simulators, all existing solutions are dedicated to a single blockchain technology and its variations. Consequently, if we want to explore what-if scenarios for different blockchains in order to chose a target platform, obviously simulators are the best option to do so. However, to facilitate such an exploration, the simulator should provide some built-in capabilities [17, 41] that can be applied to various types of blockchain systems. Currently, there is only Gauntlet that can simulate different types of blockchain systems supporting smart contracts.

## V. CONCLUSION AND PROSPECTS

Smart contracts benefit from a widespread interest because of their huge potential. However, in order to effectively build smart contracts, they need to be tested in an effective and systematic manner. In our survey, we analyzed four categories of smart contract testing approaches: using test network, using security analysis tools, using blockchain emulators with smart contracts support and using blockchain simulators with smart contract support.

Our survey shows that several VV&T solutions exist, ensure the correctness and non vulnerable patterns in smart contracts. Nevertheless, they either focus on specific security aspects, a specific blockchain and a programming language or provide limited configurations.

Besides, solutions that limit the implementation language or the environment are less general and require the user to adapt their applications which could be infeasible or very expensive. In addition, the variety of testing solutions available make it confusing and difficult as to where to start working with them.

In this way and in order to improve the VV&T process, setting up an effective method that compose between different VV&T solutions could be a better and significant approach to secure smart contracts.

---

[27] Size of Bitcoin blockchain on May 20,2021, https://www.statista.com/statistics/647523/worldwide-bitcoin-blockchain-size/, last access on 14/06/2021